\documentclass[a4paper]{jpconf}
\usepackage{graphicx}
\usepackage{iopams}
\usepackage{amssymb,latexsym}
\usepackage{amstext,amsthm}
\usepackage{amscd}
\usepackage{amsgen,amsfonts,amsbsy}

\theoremstyle{plain}
\newtheorem{theorem}{Theorem}[section]

\newtheorem{corollary}[theorem]{Corollary}

\theoremstyle{definition}

\newcommand{\Z}{\mathbb{Z}}
\newcommand{\N}{\mathbb{N}}

\newcommand{\R}{\mathbb{R}}
\newcommand{\C}{\mathbb{C}}

\newcommand{\GL}{\mathrm{GL}}
\newcommand{\Ad}{\mathrm{Ad}}
\newcommand{\Inn}{\mathrm{Int}}

\newcommand{\diag}{\mathrm{diag}}

\newcommand{\SL}{\mathrm{SL}}
\newcommand{\Sp}{\mathrm{Sp}}

\renewcommand{\sl}{\mathrm{sl}}
\renewcommand{\P}{\mathcal{P}}
\renewcommand{\H}{\mathcal{H}}
\newcommand{\A}{\mathcal{A}}

\newcommand{\be}{\begin{equation}}
\newcommand{\ee}{\end{equation}}
\newcommand{\bed}{\begin{displaymath}}
\newcommand{\eed}{\end{displaymath}}

\begin{document}

\title[A classification of finite quantum kinematics]{A classification
of finite quantum kinematics}

\author{J Tolar $^1$}
\address{
 $^1$ Department of Physics \\
 Faculty of Nuclear Sciences and  Physical  Engineering\\
 Czech Technical University in Prague\\ B\v rehov\'a 7,
 115 19 Prague 1, Czech Republic}
\eads{\mailto{jiri.tolar@fjfi.cvut.cz}}

 \begin{abstract}
Quantum mechanics in Hilbert spaces of finite dimension $N$ is
reviewed from the number theoretic point of view. For composite
numbers $N$ possible quantum kinematics are classified on the basis
of Mackey's Imprimitivity Theorem for finite Abelian groups. This
yields also a classification of finite Weyl-Heisenberg groups and
the corresponding finite quantum kinematics. Simple number theory
gets involved through the fundamental theorem describing all finite
discrete Abelian groups of order $N$ as direct products of cyclic
groups, whose orders are powers of not necessarily distinct primes
contained in the prime decomposition of $N$. The representation
theoretic approach is further compared with the algebraic approach,
where the basic object is the corresponding operator algebra. The
consideration of fine gradings of this associative algebra then
brings a fresh look on the relation between the mathematical
formalism and physical realizations of finite quantum systems.
\end{abstract}


\section{Introduction}

Looking back on our papers \cite{KoTo10, KoTo12} on symmetries of
finite Heisenberg groups, I find it appropriate --- at this
Symposium --- to return to the foundations of our approach to
quantum systems with finite-dimensional Hilbert spaces and
supplement them with the algebraic treatment.

Non-relativistic quantum mechanics of particle systems can be
divided into quantum kinematics and quantum dynamics. First the
Hilbert space and the non-commuting operators of complementary
observables, positions and momenta, are constructed. This
kinematical structure then remains the same for all possible quantum
dynamics which are determined by the respective Hamiltonians.

For systems with configuration space $\R^n$ one can define quantum
kinematics according to H. Weyl \cite{Weyl} in terms of the Weyl
system --- a projective unitary representation of the Abelian group
of translations of the corresponding classical phase space $\R^n
\times \R^n$. Using Weyl's system of unitary operators, J. von
Neumann was able to prove the uniqueness theorem stating that there
is, up to unitary equivalence, unique irreducible quantum
kinematics, commonly taken in the form of the Schr\"{o}dinger
representation. In the most general form the uniqueness theorem was
proved on the basis of G.W. Mackey's Imprimitivity Theorem. In this
mathematical generalization the configuration space $\R^n$ was
replaced by an arbitrary locally compact second countable Abelian
group $G$ \cite{Mackey}. The direct product of $G$ and its
Pontryagin dual then plays the role of the phase space.

Quantum mechanics in finite-dimensional Hilbert spaces originally
seemed to present only a nice and simple exercise in linear algebra.
During the last decades it unexpectedly became the mathematical
framework for the development of methods of quantum information
processing with numerous applications to quantum cryptography,
teleportation and quantum computing. For instance, the mathematical
notion of complementary bases lies at the heart of quantum
cryptography \cite{Gisin}.

Historically, H. Weyl --- not successful with the proof of the
uniqueness theorem --- wanted to present a simple example in $\C^n$
analogous to one-dimensional quantum particle and at the same time
exhibiting the uniqueness feature \cite{Weyl}. This example was
further developed by J. Schwinger \cite{Schwinger} with the aim to
approximate quantum mechanics of particles. He noticed the
complementary nature of position and momentum observables which are
here built in the finite Weyl-Heisenberg group generated by the
generalized Pauli matrices. The elements of this group provide a
useful basic set of quantum operators for finite quantum systems.

The special role of the finite Weyl-Heisenberg group has been
recognized also in a distant domain of mathematics
--- the classification of fine gradings of classical Lie algebras
\cite{PZ89,HPP98}. For classical Lie algebras of the type
$\sl(n,\C)$, $n=2,3,\dots$, this classification contains
--- among others --- the Pauli grading based on generalized Pauli
matrices \cite{PZ88,HPPT02}. The finite group generated by them has
been called the Pauli group --- a notion identical with the
Weyl-Heisenberg group.

In our paper \cite{StovTolar84} we succeeded to prove the uniqueness
of finite quantum kinematics using a simple version of Mackey's
Imprimitivity Theorem. In this way also a geometric interpretation
of finite quantum kinematics was obtained: it appears as quantum
mechanics where a cyclic group serves as the configuration space.
For composite dimensions $N$ the fundamental theorem on finitely
generated Abelian groups immediately leads to the classification of
all finite quantum kinematics. This classification was also
independently noted in 1995 by V.S. Varadarajan \cite{vsv}.

In this contribution our approach via representation theory is
confronted with the algebraic formulation of quantum mechanics
\cite{Emch}. For finite quantum systems, the operator algebras are
the associative complex matrix algebras $M_{N}(\C)$ \cite{Petz10}.
Inspired by Lie algebra gradings, we describe all fine gradings of
$M_{N}(\C)$ which are induced by inner automorphisms of $\GL(N,\C)$.

In section 2 we reproduce the group theoretical approach based on
representation theory of finite Abelian groups. Sections 3 and 4 are
devoted to the classification of finite quantum kinematics and some
historical remarks. In section 5 fine gradings of matrix algebras
$M_{N}(\C)$ are obtained and physical interpretation of the Pauli
gradings is offered. Section 6 concludes the paper.

\section{Simple quantum kinematics on cyclic groups}

Ordinary quantum mechanics prescribes that the mathematical
quantities representing the position and momentum should be
self-adjoint operators on the Hilbert space of the system. Their
algebraic properties constitute \textit{quantum kinematics}, while
\textit{quantum dynamics} of the system is given by the unitary
group generated by the Hamiltonian which is expressed as a function
of the position and momentum operators.

According to Mackey, quantum kinematics of a system localized on a
homogeneous space $M=G/H$ is determined by an irreducible transitive
system of imprimitivity $(\mathcal{U},\mathcal{E})$ for $G$ based on
$M$ in a Hilbert space $\H$ \cite{Mackey}. Here $\mathcal{U} = \{
U(g) \vert g \in G \}$ is a unitary representation of $G$ in $\H$
and $\mathcal{E} = \{ E(S) \vert \; S \;\; \mbox{Borel subset
of}\;\; M \}$ is a projection--valued measure in $\H$ satisfying the
covariance condition
 \begin{equation} \label{covariance}
U(g) E(S)U(g)^{-1} = E(g^{-1}.S).
 \end{equation} Given a positive integer $N \geq
2$, we assume that the configuration space $M$  is the finite set
$$ M = Z_N = \{\rho \vert \rho = 0,1,\ldots,N-1 \} $$
with additive group law modulo $N$. Since there is a natural
transitive action of $Z_N$ on itself, we may consider $M$ as a
homogeneous space of
$$ G = Z_N = \{j \vert j = 0,1,\ldots,N-1 \}, $$
realized as an additive group modulo $N$ with the action
$$ G \times M \rightarrow M :
   (j,\rho) \mapsto \rho + j (\mbox{mod} \; N)  $$
and the isotropy subgroup $H = \{0 \}.$ For the finite set $M=Z_N$
the covariance condition simplifies to
$$ U(j)E(\rho)U(j)^{-1} = E(\rho - j),    $$
where $\rho \in M$, $j \in G $, $E(\rho)=E(\{\rho\})$ and $E(S) =
\sum_{\rho \in S} E(\rho).$

Complete classification of transitive systems of imprimitivity up to
simultaneous unitary equivalence of both ${\mathcal U}$ and
${\mathcal E}$ is obtained from Mackey's Imprimitivity Theorem. Its
application to our system yields

 \begin{theorem}
 If   $({\mathcal U},{\mathcal E})$ acts irreducibly in
${\H}$, then there is, up to unitary equivalence, only one system of
imprimitivity, where:
\begin{enumerate}
   \item ${\H}$ is the Hilbert space
$\H_N = \ell^2(\Z_N)$, i.e. $\C^N$ with the inner product
 \bed
(\varphi,\psi)=\sum^{N-1}_{\rho=0}
\overline{\varphi_{\rho}}\psi_{\rho},
 \eed
where $\varphi_{\rho}$, $\psi_{\rho}$, $\rho =0,1,\ldots,N-1,$
denote the components of $\varphi$, $\psi$ in the standard basis.
    \item $ {\mathcal U}$ is the induced representation
 ${\mathcal U}= \mbox{Ind}_{H}^G\;I$ called the (right) regular representation
$$[U(j)\psi]_\rho=\psi_{\rho +j}\hskip 1cm  (j\in G).$$
Its matrix form in the standard basis is
$$(U(j))_{\rho \sigma}=\delta_{\rho+j,\sigma}.$$
  \item ${\mathcal E}$ is given by
$$[E(\rho)\psi]_{\sigma}=\delta_{\rho \sigma}
\psi_{\sigma}.$$
\end{enumerate}
\end{theorem}

This unique system of imprimitivity has a simple physical meaning.
The localization operators $E(\rho)$ are projectors on the
eigenvectors $e^{(\rho)} \in {\H}$ corresponding to the positions
$\rho = 0, 1, \ldots, N-1.$ Since the set $\{e^{(\rho)}\}$ forms the
standard basis of ${\H}$ in the above matrix realization, this
realization may be called position representation. Then in a
normalized state $\psi = (\psi_0, \ldots, \psi_{N-1})$ the
probability to measure the position $\rho$ is equal to
$$  (\psi, E(\rho) \psi) = \vert \psi_{\rho} \vert^2.$$
Unitary operators $U(j)$ act as displacement operators
$$ U(j) e^{(\rho)} = e^{(\rho - j)}.   $$
In the position representation they are given by unitary matrices
equal to the powers $U(j)=P_{N}^{j} =U(1)^j$ of the one-step cyclic
permutation matrix
 \bed
 P_{N} \equiv U(1)= \left(
\begin{array}[c]{cccccc}
0&1&0& \cdots 0&0\\
0&0&1& \cdots 0&0\\
\vdots&&&\ddots&&\\
0&0&0&....0&1\\
1&0&0&....0&0\\
\end{array}
\right). \eed

In this way finite-dimensional quantum mechanics can be viewed as
quantum mechanics on configuration spaces given by finite sets
equipped with the structure of a finite Abelian group
\cite{StovTolar84}. In the above simplest case a single cyclic group
$\Z_N$ was taken as the underlying configuration space. For given
$N\in\N$ we set $\omega_{N}:=e^{2\pi \mathrm{i}/N}\in\C$. The {\it
generalized Pauli matrices} of order $N$ are given by
 \bed
 Q_{N}:=\diag(1,\omega_{N},\omega_{N}^{2},\dots,\omega_{N}^{N-1}),
 \quad (P_{N})_{\rho,\sigma}=\delta_{\rho +1,\sigma}, \quad
\rho,\sigma\in\Z_N.
 \eed
 The subgroup of unitary matrices in $\GL(N,\C)$
generated by $Q_{N}$ and $P_{N}$,
 \bed \Pi_{N}:=\{ \omega_{N}^j Q_{N}^{k} P_{N}^{l} \vert
 j,k,l\in\{0,1,\dots,N-1 \} \} \eed
is called the {\em finite Weyl-Heisenberg group}.

The special role of the generalized Pauli matrices has been
confirmed as the cornerstone of finite-dimensional quantum mechanics
and of quantum information science. The quantum mechanical
operators, $Q_{N}$ and $P_{N}$ act in the $N$-dimensional Hilbert
space $\H_N = \ell^2(\Z_N)$. Further properties of $\Pi_{N}$ are as
follows: \textit{\begin{quote}
\begin{enumerate}
\item[(1)] The order of $\Pi_{N}$ is $N^3$.
\item[(2)] The center of $\Pi_{N}$ is
$\{\omega_{N}^\rho I_N \vert  \rho\in\{0,1,\dots,N-1 \} \}$,
 where $I_N$ is the $N \times N$ unit matrix.
\item[(3)] The commutation relation $P_{N}Q_{N}=\omega_{N}Q_{N}P_{N}$
  is equivalent with (\ref{covariance}).
\end{enumerate}
\end{quote}}

In order to formulate finite quantum kinematics in terms of the
equivalent \textit{discrete Weyl system}, we need the dual system of
unitary operators ${\mathcal V} = \{V(\rho)\}$ defined as powers of
$Q_{N}$,
 $$ V(\rho) = Q_{N}^{\rho}.$$
Then the quantum kinematics $({\mathcal U},{\mathcal E})$ can be
equivalently replaced by the discrete Weyl system $({\mathcal
U},{\mathcal V})$ satisfying
$$ U(j) V(\rho) = \omega^{j \rho}V(\rho) U(j). $$
The discrete Weyl displacement operators are defined by unitary
operators\footnote{If $N$ is odd, the factors $\omega_{N}^{j \rho
/2}$ are well-defined on $\Z_N \times \Z_N$.}
    $$ W(\rho, j) =  \omega^{j \rho /2}V(\rho) U(j)=
       \omega^{-j \rho /2} U(j)V(\rho). $$
They satisfy the composition law for a ray representation of $\Z_{N}
\times \Z_N$
$$W(\rho, j) W(\rho ', j')=  \omega^{(\rho ' j - \rho j')/2}
     W(\rho + \rho, j + j').    $$
According to Schwinger, the discrete Weyl system ${\mathcal W}$
consisting of $N^2$ operators $W(\rho,j)$ provides an operator basis
in the space of all linear operators in $\C^N$, i.e. in the full
matrix algebra $M_{N}(\C)$. The operators $W(\rho,j)$ are orthogonal
with respect to the Hilbert-Schmidt inner product
$$   \Tr(W(\rho, j) W(\rho ', j')^{*})=
  N  \delta_{\rho \rho '} \delta_{j j'}   $$
and satisfy the completeness relation
$$  \sum_{\rho,j} W(\rho, j) W(\rho ', j')^{*} =
                 N^{2} 1. $$
This important result can be summarized as follows:
\begin{theorem} \label{Schwinger}
The set of $N^2$ matrices $S(\rho, j) = Q_{N}^{\rho} P_{N}^{j}/
\sqrt{N} $, $j, \rho = 0,1,\ldots, N-1,$ constitutes a complete and
orthonormal basis for the linear space $M_{N}(\C)$ of $N \times N$
complex matrices. Any $N \times N$ complex matrix can thus be
uniquely expanded in this basis. If $N$ is odd, then the operator
basis can be taken in the form $\omega^{j\rho /2}S(\rho,j) =
W(\rho,j)/{\sqrt N}.$
\end{theorem}

\section{A classification of finite quantum kinematics}

The cyclic group
 $\mathbb{Z}_N = \left\lbrace 0,1,\ldots N-1
 \right\rbrace $
is a configuration space for $N$-dimensional quantum kinematics of a
single $N$-level system. However, the reasoning on the basis of
Mackey's Imprimitivity Theorem allows direct extension to any finite
Abelian group as configuration space because of the fundamental
theorem describing the structure of finite Abelian groups
\cite{DuFo}.
 \begin{theorem}
 Let $G$ be a finite Abelian group.
Then $G$ is isomorphic with the direct product $\Z_{N_1} \times
\cdots \times \Z_{N_f}$ of a finite number of cyclic groups for
integers $N_1, \ldots, N_f$ greater than 1, each of which is a power
of a prime, i.e. $N_{k} = p_{k}^{r_k}$, where the primes $p_k$ need
not be mutually different.
\end{theorem}
The integers $N_{k} = p_{k}^{r_k}$ are called the \textit{elementary
divisors} of $G$. Two finite Abelian groups are isomorphic if and
only if they have the same \textit{elementary divisor
decomposition}. In the special case of $G=\Z_N$ with composite $N =
p_{1}^{r_1}\ldots p_{f}^{r_f} $ and distinct primes $p_{k} > 1,$ the
unique decomposition
$$ \Z_N \cong  \Z_{N_1} \times \cdots \times \Z_{N_f}, \qquad
   N_{k} = p_{k}^{r_k}$$
is obtained by the Chinese Remainder Theorem.

Now for a general finite Abelian group as configuration space, a
transitive system of imprimitivity for $G = \Z_{N_1} \times \cdots
\times \Z_{N_f}$ based on $M = \Z_{N_1} \times \cdots \times
\Z_{N_f}$ is equivalent to the tensor product
 \be \label{tensor} {\mathcal U}
={\mathcal U}_{1}\otimes \cdots \otimes {\mathcal U}_f, \quad
{\mathcal E} ={\mathcal E}_{1}\otimes \cdots \otimes {\mathcal
E}_f
 \ee
acting in the Hilbert space
 \bed {\H} ={\H}_{1}\otimes \cdots
\otimes {\H}_f,\eed
 where the dimensions are dim ${\H}_{k}=N_k$, dim
${\H}=N_{1}\ldots N_f$. Each such system of imprimitivity
$({\mathcal U},{\mathcal E})$ is irreducible, if and only if each
$({\mathcal U}_k, {\mathcal E}_{k})$ is irreducible, hence if
irreducible, it is unique up to unitary equivalence by the
Imprimitivity Theorem.

For a given composite dimension $N$, all irreducible quantum
kinematics in the Hilbert space of dimension $N$ can be obtained
according to Theorem 3.1: just take all inequivalent choices of
elementary divisors $N_{k} = p_{k}^{r_k}$ with not necessarily
distinct primes $p_k > 1$. One can give physical interpretation to
these tensor products in the sense that the factors correspond to
the \textit{elementary building blocks or constituents} forming a
finite quantum system. Our results are summarized in
\begin{theorem}
For a given finite Abelian group $G$ there is a unique class of
unitarily equivalent, irreducible imprimitivity systems $({\mathcal
U},{\mathcal E})$ in a finite--dimensional Hilbert space ${\H}$. In
the special case $G = \Z_N$ the irreducible imprimitivity system is
unitarily equivalent to the tensor product (\ref{tensor}) with
$N_{k} = p_{k}^{r_k}$ and distinct primes $p_1,\ldots , p_f>1.$
\end{theorem}
\begin{corollary}
For a given finite Abelian group $G$ there is a unique class of
unitarily equivalent discrete Weyl systems ${\mathcal W}$ in a
finite--dimensional Hilbert space ${\H}$,
 \be {\mathcal W} \label{Weyl}
={\mathcal W}_{1} \otimes \cdots \otimes {\mathcal W}_f. \ee
\end{corollary}
The above classification conclusively shows that finite quantum
kinematics are mathematically composed of elementary constituents
which are of standard types associated with finite Abelian groups $G
= \Z_p$, $\Z_{p^2},\ldots,\Z_{p^r},\ldots$, where $p$ runs through
the set of prime numbers $>1$.

As already mentioned, among possible factorizations into elementary
constituents special role is played by the factorization associated
with the `best' prime decomposition of the dimension
$N=p_{1}^{r_1}\ldots p_{f}^{r_f} $ with mutually distinct primes
$p_1$, \dots, $p_f$. Then the discrete Weyl system of composite
dimension $N$ is equivalent to a composite system where the
constituent Weyl subsystems act in the Hilbert spaces of relatively
prime dimensions $p_{1}^{r_1}$,\dots, $p_{f}^{r_f}$.

\section{Quantum degrees of freedom taken seriously}

A historical remark is in order: the structure of the
finite--dimensional Weyl operators was thoroughly investigated by J.
Schwinger. He noted in particular that, if the dimension $N$ is a
composite number which can be decomposed as a product
$N=N_{1}N_{2}$, where the positive integers $N_1$, $N_2$ $>1$ are
relatively prime, then all the Weyl operators in dimension $N$ can
be simultaneously factorized in tensor products of the Weyl
operators in the dimensions $N_1$ and $N_2$. J. Schwinger then
states ( \cite{Schwinger}, pp. 578--579):
 \begin{quote}
\textit{``The continuation of the factorization terminates in
$$       N = \prod_{k=1}^{f} \nu_k,  $$
where $f$ is the total number of prime factors in $N$, including
repetitions. We call this characteristic property of $N$ the number
of degrees of freedom for a system possessing $N$ states.''} And
further, \textit{``each degree of freedom is classified by the value
of the prime integer $\nu = 2,3,5,\ldots \infty$.''}
 \end{quote}

However, our application of Theorem 3.1 leads to the building blocks
$\Z_{p^r}$ with $r\geq 1$ as constituent configuration spaces. This
fact was independently noted by V.S. Varadarajan in his paper
\cite{vsv} devoted to the memory of J. Schwinger:
\textit{``Curiously, Schwinger missed the systems associated to the
indecomposable groups $Z_N$ where $N$ is a prime power $p^r$, $r\geq
2$ being an integer.''} And in another paper: \textit{''In this way
he arrived at the principle that the Weyl systems associated to
$\Z_p$ where $p$ runs over all the primes are the building blocks.
Curiously this enumeration is incomplete and one has to include the
cases with $\Z_{p^r}$ where $p$ is as before a prime but $r$ is any
integer $\geq 1$.''}

Thus the elementary building blocks of finite quantum kinematics are
given by the elementary divisor decomposition. In mathematics also
exists equivalent \textit{invariant factor decomposition}
\cite{DuFo}. In that approach a finite Abelian group $G$ is uniquely
determined by an ordered finite list of integers $n_1\geq n_2 \geq
\dots \geq n_s$ greater than 1 determining the invariant factors
$\Z_{n_i}$ of $G$ such that $n_{i+1}$ divides $n_{i}$ and
$N=n_{1}n_{2}\dots n_{s}$. For instance, if $N=180$, the full list
of non-isomorphic Abelian groups of order $180$ consists of
$\Z_{180}$, $\Z_{2}\times \Z_{90}$, $\Z_{3}\times \Z_{60}$,
$\Z_{6}\times \Z_{30}$. However, to exhibit the elementary building
blocks of the corresponding composite system, the elementary divisor
decomposition is indispensable. In the above example all
inequivalent choices of elementary divisors $N_{k} = p_{k}^{r_k}$
with not necessarily distinct primes $p_k > 1$ are $2^{2}.3^{2}.5$,
$2.2.3^{2}.5$, $2^{2}.3.3.5$, $2.2.3.3.5$.

The task of enumerating all finite quantum kinematics in dimension
$N$ then amounts to the determination of all finite Abelian groups
of order $N$. It starts with the factorization of
$N=p_{1}^{r_1}\ldots p_{f}^{r_f} $ with mutually distinct primes
$p_1$, \dots, $p_f$. First one finds all permissible lists for
groups of orders $p_{i}^{r_i}$ for each $i$. For a prime power
$p_{i}^{r_i}$ the problem of determining all permissible lists is
equivalent to finding all \textit{partitions of the exponent} $r_i$,
and does not depend on $p_{i}$. Recall that the number of partitions
of a natural number $r$ is called \textit{Bell's number} $B(r)$.
Then the total number of groups of order $N$ is equal to the product
of Bell's numbers $B(r_1)B(r_2)\ldots B(r_f)$.

In physics, the dimensions of constituent Hilbert spaces are
primarily fixed by the numbers of levels of physical subsystems. It
follows that $G$ is isomorphic to a direct product of cyclic groups
of the respective orders. Whenever this is the case, in order to be
able to work with the elementary building blocks of the
corresponding quantum kinematics, we should find the elementary
divisors. For example, if $G=\Z_{6}\times\Z_{15}$, we factor $6=2.3$
and $15=3.5$. Then the elementary divisor decomposition of $G$ is
$\Z_{2}\times\Z_{3}\times\Z_{3}\times\Z_{5}$.

\section{Algebraic approach}

In quantum information theory as in algebraic quantum theory, the
operators of a finite quantum system belong to the full matrix
algebra $M_{N}(\C)$ of $N\times N$ complex matrices. In this
contribution we started with finite Abelian groups and, via
representation theory expressed by Mackey's systems of
imprimitivity, arrived at the system of $N^2$ unitary operators
forming a complete orthonormal basis of the linear space $M_{N}(\C)$
with the Hilbert-Schmidt inner product.

Now we shall pass the opposite journey starting from the operator
algebra $M_{N}(\C)$ and equip it as associative algebra with
additional structure of a fine grading induced by commuting inner
automorphisms of $\GL(N,\C)$.

A \textit{grading} of an associative algebra $\A$ is defined as a
direct sum decomposition of $A$ as a vector space
 \begin{equation} \label{grading}
  \Gamma \; : \quad  \A = \bigoplus_{\alpha} \A_{\alpha}
 \end{equation}
satisfying the property
 $$
  x \in \A_{\alpha}, \quad y \in \A_{\beta}\quad \Rightarrow \quad xy \in
  \A_{\gamma}.
 $$
Similarly as for Lie algebras \cite{PZ89}, the linear subspaces
$\A_{\alpha}$ can be determined as eigenspaces of automorphisms of
the associative algebra $\A$. For an automorphism $g$ of $\A$,
$g(xy) = g(x) g(y)$ holds for all $x,y \in \A$. Now if
$g(x)=\lambda_{\alpha} x$ defines the subspace $\A_{\alpha}$ and
$g(y)=\lambda_{\beta} y$ defines the subspace $\A_{\beta}$, then
 $$
 g(xy) = g(x) g(y) = \lambda_{\alpha}\lambda_{\beta} xy
 $$
defines the subspace $\A_{\gamma}$ with
$\lambda_{\gamma}=\lambda_{\alpha}\lambda_{\beta}$. In this way $g$
induces the grading decomposition $\A =
\bigoplus_{\alpha}\mbox{Ker}(g-\lambda_{\alpha})$. Further commuting
automorphisms may refine the grading. If $gh=hg$, we have
 $$
 g(h(x)) = h(g(x)) = h(\lambda_{\alpha}x) = \lambda_{\alpha}h(x)
 $$
implying that the grading decomposition using both $g$ and $h$ may
lead to a refinement. By extending the set of commuting
automorphisms one arrives at so called \textit{fine gradings} where
the grading subspaces have the lowest possible dimension.

For the associative algebra $M_{N}(\C)$ we shall look for fine
gradings induced by the inner automorphisms. For $M\in\GL(N,\C)$ we
denote $\Ad_{M}\in \Inn(M_{N}(\C))$ be the \textit{inner
automorphism} of $M_{N}(\C)$ induced by operator $M\in \GL(N,\C)$,
i.e.
 \bed \Ad_{M}(X)=MXM^{-1}\quad \textrm{for}\quad X\in M_{N}(\C).
 \eed
The relevant properties of $\Ad_{M}$ are: for $M,N\in \GL(N,\C)$
 \textit{\begin{enumerate}
\item[(i)] $\Ad_M \Ad_N=\Ad_{MN}$.
\item[(ii)] $(\Ad_M)^{-1}=\Ad_{M^{-1}}$.
\item[(iii)] $\Ad_{M}=\Ad_{N}$ if and only if  there is a constant
$0\neq\alpha\in\C$ such that $M=\alpha N$.
\end{enumerate}}

Since the commuting inner automorphisms form an Abelian subgroup of
$\Inn(M_{N}(\C))$, \textit{fine gradings} of $M_{N}(\C)$ can be
obtained using the \textit{maximal Abelian groups of diagonalizable
automorphisms} -- as subgroups of $\Inn(M_{N}(\C))$ -- which have
been called the MAD-\textit{groups} \cite{HPP98}. Thus we are
looking for fine gradings which are induced via diagonalizable
elements of maximal Abelian subgroups of $\Inn(M_{N}(\C))$.

The MAD-groups of inner automorphisms of $M_{N}(\C)$ can be derived
by following \cite{HPP98}. The result is straightforward and is
summarized in the following theorem:
\begin{theorem}
Any MAD-group contained in $\Inn(M_{N}(\C))$ is conjugated to one
and only one of the groups of the form
 \begin{equation} \label{MAD}
  \P_{N_1} \otimes \P_{N_2} \otimes \ldots \otimes \P_{N_f} \otimes
  D(m),
 \end{equation}
where $N_i=p_{i}^{r_i}$ are powers of primes, $N=N_{1}N_{2}\dots
N_{f} \ m$ and $D(m)$ is the image in $\Inn(M_{N}(\C))$ of the group
of $m \times m$ complex diagonal matrices under the adjoint action.
\end{theorem}
Here $\P_N$ is defined as the group
 $$ \P_N=\{\Ad_{Q_N^i P_N^j} \vert (i,j)\in \Z_N \times \Z_N \}.$$
It is an Abelian subgroup of $\Inn(M_{N}(\C))$ and is generated by
two commuting automorphisms $\Ad_{Q_N}$, $\Ad_{P_N}$, each of order
$N$. A geometric view is sometimes useful that $\P_N$ is isomorphic
to the \textit{quantum phase space} identified with the Abelian
group $\Z_N \times \Z_N$.

For illustration, we give a list of MAD-groups in low dimensions:
\begin{itemize}
 \item $n=2$: $\P_{2}\otimes D(1)$, $D(2)$
 \item $n=3$: $\P_{3}\otimes D(1)$, $D(3)$
 \item $n=4$: $\P_{4}\otimes D(1)$,
        $\P_{2}\otimes \P_{2}\otimes D(1)$,
        $\P_{2}\otimes D(2)$, $D(4)$
 \item $n=5$: $\P_{5}\otimes D(1)$, $D(5)$
 \item $n=6$: $\P_{3}\otimes \P_{2}\otimes D(1)$, $\P_{3}\otimes D(2)$,
           $\P_{2}\otimes D(3)$, $D(6)$
 \item $n=7$: $\P_{7}\otimes D(1)$, $D(7)$
 \item $n=8$: $\P_{8}\otimes D(1)$, $\P_{4}\otimes \P_{2}\otimes D(1)$,
    $\P_{2}\otimes \P_{2}\otimes \P_{2}\otimes D(1)$,
    $\P_{4}\otimes D(2)$, $\P_{2}\otimes \P_{2}\otimes D(2)$,
    $\P_{2}\otimes D(4)$, $D(4)$
\end{itemize}

A part of the obtained MAD-groups containing the trivial diagonal
subgroup $D(1)$ induce exactly all our Pauli decompositions
(\ref{tensor}), (\ref{Weyl}). However, there are still fine gradings
induced by $D(m)$, $m=2,3 \dots$. They include partial or complete
decompositions which have the form of the Cartan root decompositions
of Lie algebras $\sl(m,\C)$ extended by the unit matrix. They
contain the Abelian Cartan subalgebra of dimension $m-1$, the unit
matrix and one-dimensional root subspaces spanned by nilpotent
matrices. Concerning physical interpretation of these Cartan parts
of the decompositions, one can speculate that they may reflect extra
degrees of freedom corresponding to some internal symmetries
\cite{Kibler}. Leaving these decompositions aside, we are left with
the \textit{Pauli gradings} which decompose $M_{N}(\C)$ in direct
sums of $N^2$ one-dimensional subspaces. For these Pauli
decompositions with given $N$ one should realize that $M_{N}(\C)$
encompasses all operators of any quantum system with $N$-dimensional
Hilbert space.

In this way we have got a new view on the relation between the
general mathematical formalism and physical realizations of finite
quantum systems. The apparent contradiction that $M_{N}(\C)$
represents any $N$-dimensional quantum system and at the same time
there is a multitude of inequivalent quantum kinematics for given
$N$, is simply resolved: from the physical point of view the same
algebra $M_{N}(\C)$ is the operator algebra not only for a single
$N$-level system, but also for all other members of the set of
inequivalent quantum kinematics for this $N$.\footnote{Note that
from the mathematical point of view $M_{N_1}(\C) \otimes \cdots
\otimes M_{N_f}(\C)$ is isomorphic with $M_{N}(\C)$.} They just
correspond to different physical realizations of composite quantum
systems. Of course, each such system has its preferred quantum
operators (\ref{Weyl}).

\section{Conclusions}

In our study unexpectedly rich structures were obtained from
number-theoretic properties connected with prime decompositions of
numbers $N$. Our studies may also shed light on a long-standing
unsolved problem related to complementary observables in
finite-dimensional quantum mechanics. There the notion of
complementarity of observables $A, B$ with non-degenerate
eigenvalues is equivalently reformulated in terms of their
eigenvectors forming mutually unbiased bases: if the system is
prepared in any eigenstate of $A$, then the transition probabilities
to all eigenstates of the complementary observable $B$ are the same
(equal to $1/N$). It is known that the maximal set of mutually
unbiased bases contains at most $N+1$ bases and that this maximal
number is attained for $N$ prime or a power of a prime. For
composite numbers $N$ the maximal number of mutually unbiased bases
is unknown. The needed bases can be constructed as common
eigenvectors of suitable subsets formed by commuting Pauli
operators. In some cases the decomposition of the set of all Pauli
operators into subsets of commuting operators can be reflected in
the finite geometry \cite{RosuPlanatSaniga04}. The study of mutually
unbiased bases may have implications for quantum information and
communication science, since mutually unbiased bases are
indispensable ingredients of quantum key distribution protocols
\cite{Gisin}.

There exist numerous studies of various aspects of the finite
Weyl-Heisenberg group over finite fields for Hilbert spaces of prime
or prime power dimensions, e.g. \cite{Balian,Fivel,Neuhauser,
SulcTolar07,Vourdas07}. However, our main motivation to study finite
quantum kinematics not in prime or prime power dimensions but for
arbitrary dimensions stems from our previous research where we
obtained results valid for arbitrary dimensions \cite{HPPT02,
StovTolar84,TolarChadz}. There are also other papers supporting our
motivation \cite{VourdasBanderier, Digernes,DigernesVV}.

Finite quantum systems are basic constituents of quantum information
processing. Except single $2$- and $d$-level quantum systems
--- qubits and qudits --- many authors pay their attention
to finitely composed systems, where the basic operators are formed
by tensor products. Multiple qubits with Hilbert spaces
$\C^{2}\otimes \ldots \otimes \C^{2}$ are routinely employed in
quantum algorithms, while multiple qudits with Hilbert spaces
$\C^{d}\otimes \ldots \otimes \C^{d}$ may be interesting for quantum
error-correction codes and for multipartite communication. What
about general systems? They are sometimes called `mixtures of
multiple qudits'. We have shown the general classification of finite
quantum kinematics and their physical interpretation: for given
dimension there exists a broad variety of finitely composed distinct
quantum kinematics involving inequivalent tensorial factorizations
into elementary constituents. It is remarkable that in spite of the
boom of quantum information science the community of quantum
information and communication has not noticed this general
classification up to now. Only the visible aspect of the
finite-dimensional operator formalism is perceived in terms of
examples constructed from generalized Pauli matrices in an {\em ad
hoc} manner. In this way the underlying general structure remains
hidden. In our approach we explicitly exhibit the elementary quantum
degrees of freedom, since their relevance has not been emphasized in
most of the literature on finite quantum systems.

It is clear that automorphisms or symmetries of the finite
Weyl-Heisenberg group play very important role in the investigation
of Lie algebras on the one hand \cite{HPPT02,HNPT06,PST06} and of
quantum mechanics in finite dimensions on the other
\cite{Vourdas,SulcTolar07}. These symmetries find proper expression
in the notion of the quotient group of a certain normalizer
\cite{PZ89}. The groups of symmetries given by inner automorphisms
were described in \cite{HPPT02} as isomorphic to $\SL(2,\Z_N)$ for
arbitrary $N \in \N$ and as $\Sp(4,\Z_p)$ for $n=p^2$, $p$ prime in
\cite{PST06} (see also \cite{Han10}). For the complete description
of these symmetries --- in quantum information conventionally called
Clifford groups \cite{Gottesman} --- we refer to our papers
\cite{KoTo10, KoTo12}.

\section*{Acknowledgements}
The author is grateful to M. Havl\'{i}\v{c}ek, M. Korbel\'a\v{r} and
P. Novotn\'y for fruitful discussions on the subject of the paper.
Support by the Ministry of Education of Czech Republic from the
projects MSM6840770039, LC06002 and the research plan RVO:68407700
is acknowledged.

\end{document}